\documentclass[jkps,preprint,fleqn,showpacs,showkeys]{revtex4}
\usepackage{graphicx}
\usepackage{amssymb}
\usepackage{amsmath}
\usepackage{bm}
\begin{document}
\setcounter{page}{0}
\title[]{RRR Characteristics for SRF Cavities}
\author{Yoochul \surname{Jung}}
\email{sulsiin@ibs.re.kr}
\thanks{Fax: +82-42-878-8866}
\author{Myungook \surname{Hyun}}
\author{Mijoung \surname{Joung}}

\affiliation{RISP, Institute for Basic Science, Daejon
305-811}

\date[]{Received 7 November 2014}

\begin{abstract}
The first heavy ion accelerator is being constructed by the rare isotope science project (RISP)
launched by the Institute of Basic Science (IBS) in South Korea. 
Four different types of superconducting cavities were designed, and prototypes were fabricated
such as a quarter wave resonator (QWR), a half wave resonator (HWR) and a single spoke resonator (SSR).
One of the critical factors determining performances of the superconducting cavities is a residual resistance ratio (RRR).
The RRR values essentially represent how much niobium is pure and how fast niobium can transmit heat as well.
In general, the RRR degrades during electron beam welding due to the impurity incorporation. 
Thus it is important to maintain RRR above a certain value at which a niobium cavity shows target performance. 
In this study, RRR degradation related with electron beam welding conditions, 
for example, welding power, welding speed, and vacuum level will be discussed. 
\end{abstract}



\maketitle

\section{INTRODUCTION}

Surface resistance is very important in the operation of superconducting cavity \cite{HasanPadamsee1,HasanPadamsee2}. 
As one can see from equation (1),
surface resistance is the function of a temperature dependent term 
and a temperature independent one. Temperature dependent resistance simply decreases
as temperature goes down in a metal since the number of phonons decreases according
to Boltzmann factor \cite{Kittel}.  
\begin{equation}
\ R_{surface} = \ R_{{temp-dependent}, {BCS}} ~ + \ R_{{temp-independent}, {Residual}}
\label{eq:two}
\end{equation}

Practically, the operating temperature of a superconducting cavity is usually determined
in such a way to achieve possibly low surface resistance as long as an economical issue allows.
In other words, the operating temperature cannot be infinitely low 
since lowering temperature always demands substantial cost. 
For the Raon accelerator, a quarter wave resonator (QWR) cavity will be operated at 4K
while the rest cavities will be operated at 2K \cite{JKPS14Jeon}. 

Unlike the temperature dependent resistance, the temperature independent resistance,
which is called a residual resistance, is due to various factors. 
It has been reported that defects acting scattering centers contribute
to the residual resistance such as inclusions, voids, dislocations with a crystal lattice itself \cite{Barrett}.
Thus, lowering residual resistance is essential to achieve low surface resistance. 
Basically, the more pure material is, the lower residual resistance one can obtain.  
Since the Residual Resistance Ratio (RRR) is defined as the ratio of a resistance at 300K
to a residual resistance (resistance at or just above critical temperature,
$ T_c$ of Nb $\sim$ 9.2K) \cite{Research11Splett}, one can have high RRR with the more pure material.  

Fabrication process of superconducting cavities is very complicated
since it requires mechanical (pressing), electrical (electron beam welding),
and chemical (polishing) process \cite{PhysRev11Chen,IPAC14Jung}. 
Among these steps, the electrical process (e-beam welding) is of our interest in this paper. 
Generally, the RRR degrades during e-beam welding due to the introduction of impurities 
from surroundings such as oxygen, nitrogen, and hydrogen \cite{AIP03Singer}.
Although previous studies reported that a vacuum level of  a chamber is the most critical,
but still e-beam welding conditions, such as e-beam power and welding speed,
are considered important as much as the vacuum level
because they make a great effect on the heat affected zone (HAZ) in a welding part \cite{IEEE09Champion}.

In this paper, we report RRR degradation occurred during e-beam welding of niobium samples. 
We measured RRR with different conditions; vacuum level, e-beam power, and welding speed.  
Not only we compared RRR degradation, but also we analyzed a heat affected zone (HAZ).
Thus, we finally report how RRR degradation and heat affected zone were affected by different e-beam welding conditions.

\section{Experiments and discussion}
\begin{figure}
\centering
\includegraphics[width=16.0cm]{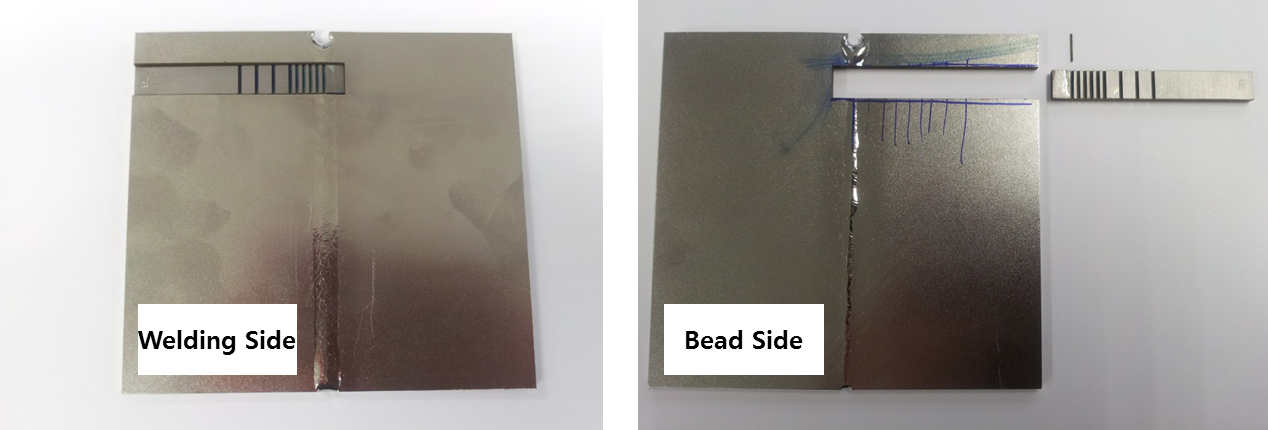}
\caption {Welded Nb sheet for RRR sample preparation by company B. Nb sheet of  $200 \times 200$ mm$^{2}$ 
was butt-welded by joining two Nb sheet of  $100 \times 200$ mm$^{2}$. Weld side (Left), Bead side (Right).
}\label{p1}
\end{figure}

\begin{figure}
\centering
\includegraphics[width=16.0cm]{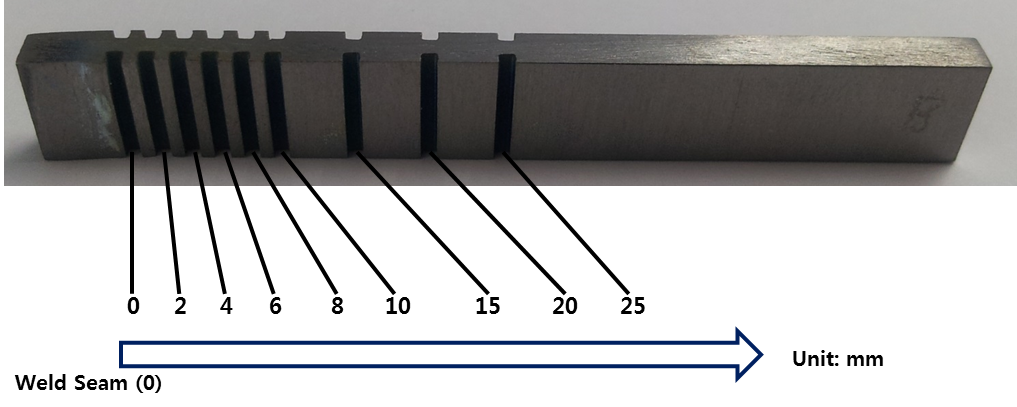}
\caption{RRR sample preparation. Eighteen samples were cut out of the welded Nb sheet
(nine from the weld side and another nine from the bead side)
with the coordination from the weld seam (zero) to the end point (25mm) perpendicular
to a welding line. Each Nb sample was cut by Electrical Discharge Machining (EDM)
with the size of is $0.5 \times 0.5 \times 9$ mm$^{3}$ for RRR measurement.
}\label{p2}
\end{figure}

A RRR 300 grade niobium sheet was from ATI, Wah Chang Inc. (Albany, USA.)
with the size of $3 \times 635 \times 1200$ mm$^{3}$ ($thickness \times width \times length$).
The chemical composition and the grain size of Nb sheet satisfies each ASTM $\#5$ and ASTM $\#4$.
Samples for experiments were prepared from two different companies (A \& B) and the sequences are as follows.
First, two companies cut the Nb sheet of 3 mm in thickness into two different sets of width and length
for butt welding \cite{PAC03Jiang}: two Nb pieces of $50\times 100$ mm$^{2}$ 
by company A and two Nb pieces of $100 \times 200$ mm$^{2}$ by company B.
Thus, the company A prepared the welded Nb sheet of  $100 \times 100$ mm$^{3}$,
and the company B prepared the welded Nb sheet of $200 \times 200$ mm$^{2}$.
Then, RRR samples were cut out of these welded Nb sheets perpendicular to a welding line. 
Each Nb sample was cut by Electrical Discharge Machining (EDM)
\cite{International07Abbas} with the size of $0.5 \times 0.5 \times 9$ mm$^{3}$.
Fig. \ref{p1} shows how the pieces of Nb sheet were welded,
and samples were cut out from the welded Nb sheet by the company B. 
Also, Fig. \ref{p2} shows the layout of cut samples, all samples were perpendicular to the welding line, 
and the distance from the weld seam increased from left to right (0, 2 ,4, 6, 8, 10, and 15, 20, 25 mm).
As shown in Fig. \ref{p2}, in order to compare differences in RRR between the welding part and the bead part, 
two types of samples were cut from the welding side and the bead side. 
Since nine samples were cut out from each side,
eighteen samples in total were prepared from both sides of one welded Nb sheet. 
Welding conditions of two companies are summarized in Table~ \ref{t1}.

In this study, we defined the RRR value as the ratio of the Nb resistance at 300K to the 
Nb resistance at the temperature just above critical temperature ($T_{c,Nb}$ $\sim$ 9.2K)
where Nb still is in the normal conducting state. That temperature ranges from 9.5K to 10K.
The input current was 5 mA, and RRR measurement was performed with no magnetic field application. 
Since the generated power in the sample ($\sim$ 10$^{-10}$ Watt) due to this level of input current is negligible
compared to the cooling power of sample stage in PPMS ($\sim$ 10$^{-3}$ Watt), the 5mA input current level dose not vary the resistance of the sample
while measurement.
The time for stabilizing each temperature step was between 100 $\sim$ 200 sec.
The temperature increments were 0.1K up to 15K, 1K from 16K to 30K, and 5K from 35K to 300K.
RRR experiments were carried out with Physical Property Measurement System (PPMS)
operated by the Korea Basic Science Institute (KBSI). 

\begin{figure}
\centering
\includegraphics[width=16.0cm]{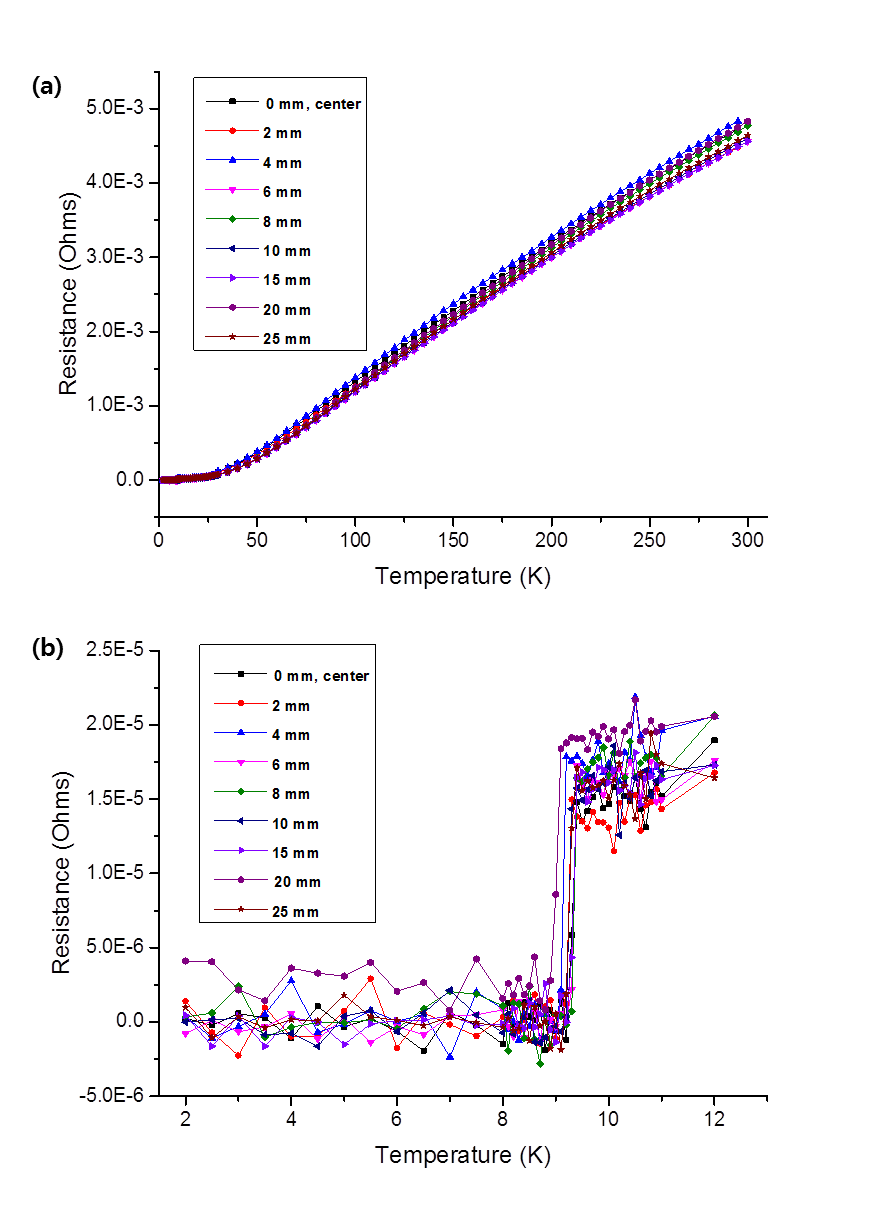}
\caption{Temperature versus resistance of the samples (company A, weld side), (a) temperature dependence of resistance from 300 K to 2 K
(b) temperature dependence of resistance from 12 K to 2 K
}\label{p3}
\end{figure}

\begin{table}
\caption{Welding Conditions for Sample Preparation}
\begin{ruledtabular}
\begin{tabular}{cccccc}
 & Voltage (kV) & Current (mA) & Power (kW)& Speed (mm/s) & Vacuum (Torr)\\
\colrule
 Company A & 60 - 120 & 30 - 60 & Max. 7.2 & 3 - 20 & $5.4\times 10^{-6}$  \\
 Company B & 60 - 150 & 30 - 60 & Max. 9 & 3 - 20 & $2   \times 10^{-5}$ \\
\end{tabular}
\end{ruledtabular}
\label{t1}
\end{table}

\begin{figure}
\centering
\includegraphics[width=15cm]{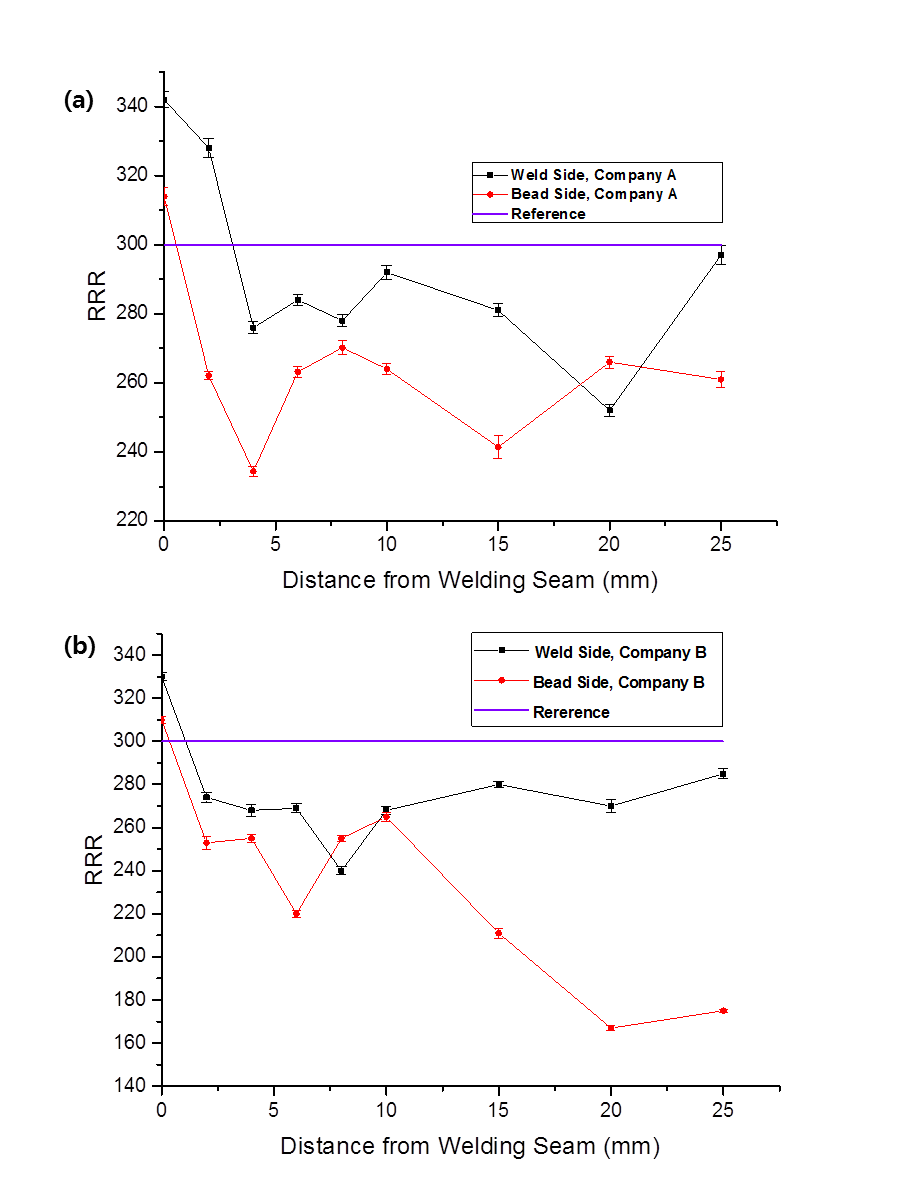}
\caption {RRR vs. Distance from the weld seam. (a) Company A (b) Company B.
Lines of square-symbol and lines of circle-symbol represent RRR each from the weld side and the bead side.
}\label{p4}
\end{figure}

\begin{figure}
\centering
\includegraphics[width=15cm]{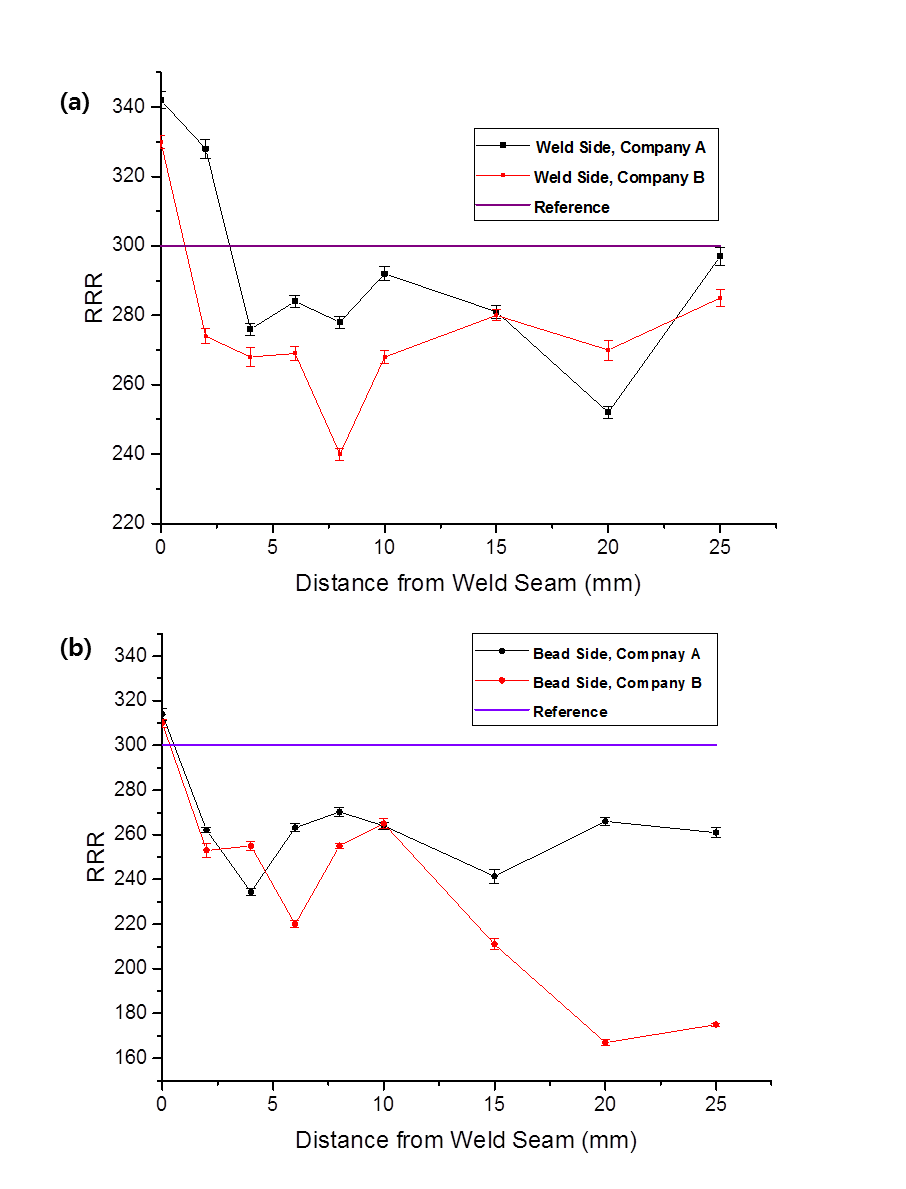}
\caption {RRR comparison, Weld Sides vs. Bead Sides (a) Weld sides of company A and B,
(b) Bead sides of company A and B. Lines of square-symbol and lines of circle-symbol
 represent RRR each from the weld sides and the bead sides.
}\label{p5}
\end{figure}

\begin{figure}
\centering
\includegraphics[width=16.0cm]{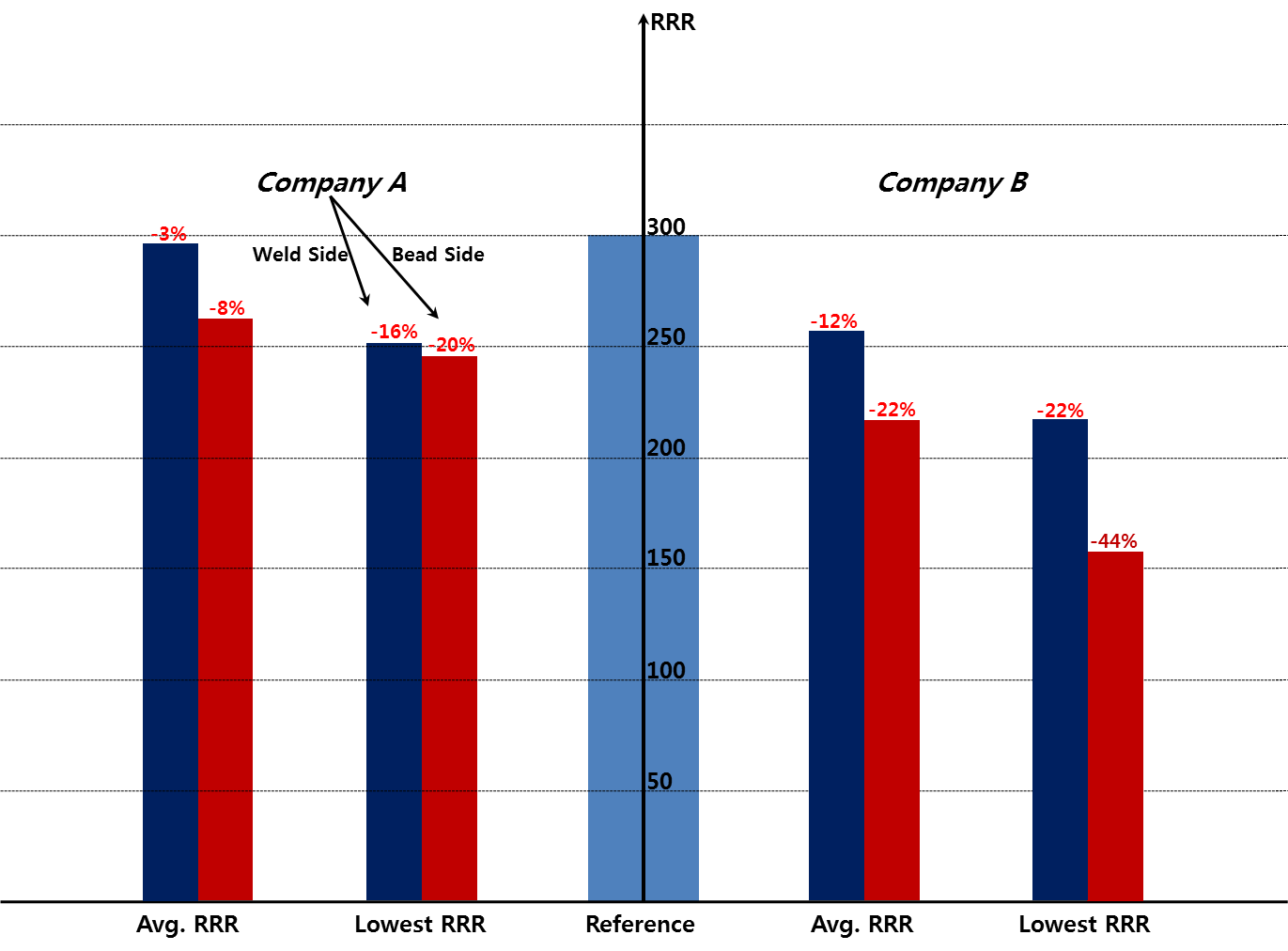}
\caption{RRR degradation rates for all samples. Degradation based on Avg. RRR and lowest RRR
are shown with regard to the reference sample of 300 RRR grade Nb. 
The left blue bar represents the degradation rate of the weld side,
and the right red bar represents the degradation rate of the bead side. 
}\label{p6}
\end{figure}

The temperature dependent resistances of the weld side samples supplied by the company A are shown in Fig. \ref{p3}.
Figure \ref{p4} shows RRR results obtained from company A and B as the function of the distance 
from the weld seam. The standard deviations of every single resistance point for all cases are additionally represented with error bars.
The single RRR point was obtained by using the average values of and at least 5 different resistance points between 10 K and 9.5 K.
Square-symbol lines and circle-symbol lines represent RRR results each from the weld side
and the bead side. As one can see, in (a), (b) of  the Fig. \ref{p4}, 
the weld side showed higher RRR than the bead side for both companies. 
In order to compare two companies regarding same sides, Fig. \ref{p4} are redrawn in Fig. \ref{p5}. 
From the Fig. \ref{p5}, company A showed higher RRR than company B in both the weld sides and bead sides.
Average RRR, the lowest RRR, and the degradation rate of RRR were calculated for both companies and listed in Table ~\ref{t2}, and Table ~\ref{t3}.
The degradation rates of RRR were calculated based on the average RRR and the lowest RRR value. 
For the comprehensive comparison of RRR data, all above figures are redrawn as Fig. \ref{p6}.
The rates of RRR degradation for all samples versus the reference sample of 300 RRR grade Nb are shown in Fig. \ref{p6}.

\begin{table}
\caption{RRR Data of Company A}
\begin{ruledtabular}
\begin{tabular}{ccccc}
 & Avg. RRR & Lowest RRR & Degradation(Avg. RRR) & Degradation(Lowest RRR)\\
\colrule
Weld Side & 292 & 252 & 3$\%$ & 16$\%$ \\
Bead Side & 276 & 240 & 8$\%$ & 20$\%$ \\
\end{tabular}
\end{ruledtabular}
\label{t2}
\end{table}

\begin{table}
\caption{RRR Data of Company B}
\begin{ruledtabular}
\begin{tabular}{ccccc}
 & Avg. RRR & Lowest RRR & Degradation(Avg. RRR) & Degradation(Lowest RRR)\\
\colrule
Weld Side & 264 & 234 & 12$\%$ & 22$\%$  \\
Bead Side & 235 & 167 & 22$\%$ & 44$\%$ \\
\end{tabular}
\end{ruledtabular}
\label{t3}
\end{table}

From the Fig. \ref{p6}, RRR degradation of the samples from company A occurred less than company B. 
This is in good agreement with previous studies \cite{AIP03Singer,RF03Bauer}
since RRR degradation is greatly affected by a vacuum level.
As shown in Table ~\ref{t1}, the vacuum levels of two companies each are $5.4\times 10^{-6}$ torr for company A,
and $2\times 10^{-5}$ torr for company B. Therefore, we could confirm that samples from company A showed higher
RRR values (less degradation) than company B as the average RRR (292 vs. 264), and the lowest RRR (253 vs. 234).
The worst degradation rates based on the lowest points were each 20\% and 44 \% corresponding to company A and B. 
In fact, RISP set the minimum RRR value as 275, so only less than 8\% of degradation is allowable in RAON project.
Although the degradation rate of samples from company A satisfied this criteria as 3\% and 8\% in the weld side
(these was based on the average RRR), the rest degradation rates did not satisfy the criteria.
We might think this was due to the vacuum level, that is to say, the vacuum level gauge did not read precisely
the real vacuum level in the e-beam chamber, thus the real vacuum level during the welding did not reach
$5.4\times 10^{-6}$ torr.
So we should perform more experiments to confirm this result.

Interesting results were that the RRR degradation in the weld side where the e-beam welding directly occurred
was less than in the bead side for all samples regardless of companies.
According to previous studies \cite{Met09Choi,Mat00Koethe}, the part where the e-beam welding directly occurred
showed high RRR value since this part experienced purification due to the "melting"
just like the Zone Melting method used in such as a Nb purification
and a semiconductor purification \cite{Barrett,Pfann}.
This result can be explained from the fact that a liquid phase can hold more solute atoms (impurities)
than a solid phase, and the melting the solid into the liquid phase drives impurities from solid to liquid.
Then, impurities in the liquid are driven into the solid quickly on cooling since the diffusivity of impurities
in the liquid can be assumed evenly faster than in the solid. 
Thus, it makes adjacent solid part has more impurities \cite{IEEE09Champion}.
Therefore, the driven effect of impurities caused the weld side to be more purified
than the bead side although these two parts experienced the melting event during the e-beam welding.

Another result to be discussed here is a heat affected zone (HAZ).
The RRR degradation occurs seriously in this region 
because grains in this region are greatly altered by a heat introduction
depending on the e-beam power and the welding speed
while RRR recovers as much as bulk material outside this region.
Therefore, controlling (narrowing) the HAZ is an important issue for
optimizing e-beam welding to keep RRR degradation above the target value.
From the W. Singer's work \cite{AIP03Singer},
the HAZ was around 10mm from the weld seam.
By looking at Fig. \ref{p3} and Fig. \ref{p4}, the HAZ was around 15mm where RRR started to recover
as high as bulk's value for both companies. 
In particular, the RRR of the bead side from company B was still far lower than the reference sample
up to 25mm, which means the HAZ expanded into the bulk level. 
One possible explanation is because too much heat was introduced in the welding part,
and this heat did not spread out quickly through the whole sample during e-beam welding. 
In fact, Nb has a low thermal conductivity, for example,
the thermal conductivity of copper is larger than that of niobium by one order.  
Therefore, the welded part of Nb did not have enough time to dissipate overheat quickly into the whole sample
due to the low thermal conductivity.
Consequently, the development of the optimized welding condition
including the welding power and the welding speed should be established in order to achieve  
good HAZ (narrow HAZ).
As a next step, we need more experiments to analyze quantitatively how much heat was generated
in the welding zone and how much grains were affected with the function of welding power and speed
by using high quality optical microscope or SEM.

\section{CONCLUSIONS}

The RRR measurements with the 300 RRR grade Nb samples supplied from two companies
of different welding conditions were carried out.
We confirmed that the vacuum level was critical factor to avoid RRR degradation
since the degradation rate of company A having  $5.4\times 10^{-6}$ torr
was lower than that of company B having  $2\times 10^{-5}$ torr.
Also, we found that the degradation of the weld side where the melting directly occurred 
showed lower degradation than the bead side. In addition, we found that the HAZ could be
varied by controlling the welding condition, which was the heat introduction. 
The HAZ expanded deeper into the bulk when the heat was introduced too much in the welding zone.

\begin{acknowledgments}
We thank KBSI for performing RRR tests for this study. This work was supported by the Rare Isotope Science Project which is funded by the Ministry of Science, ICT and Future Planning (MSIP) and the National Research Foundation (NRF) of the Republic of Korea under Contract 2011-0032011.
\end{acknowledgments}


\begin{references}
\bibitem{HasanPadamsee1} H. Padamsee, J. Knobloch, T. Hays, {\em RF Superconductivity for Accelerators} (Wiley-VCH Verlag GmbH $\&$ Co. KGaA, Germany, 2008).
\bibitem{HasanPadamsee2} H. Padamsee, {\em RF Superconductivity, Science, Technolgoy, and Applications} (Wiley-VCH Verlag GmbH $\&$ Co. KGaA, Germany, 2009).
\bibitem{Kittel} C. Kittel, {\em Introduction to Solid State Physics} (John Wiley $\&$ Son, Inc., United States, 1996).
\bibitem{JKPS14Jeon} D. Jeon {\it et al.}, J. Kor. Phys. Soc. {\bf 65}, 1010 (2014).
\bibitem{Barrett} C. R. Barrett, {\em The Principles of Engineering Materials} (Prentice Hall, United States, 1973).
\bibitem{Research11Splett} J. D. Splett, D. F. Vecchia, L. F. Goodrich, J. Research National Inst. Standards and Tech, {\bf 116}, 489 (2011).
\bibitem{PhysRev11Chen} C. Xu, H. Tian, C. E. Reece, M. J. Kelly, Phys. Rev. Special Topics - Accelerators and Beams, {\bf 15}, 043502 (2012).
\bibitem{IPAC14Jung} Y. Jung, H. J. Kim, H. H. Lee, H. C. Yang, {\em Proc. 5th Int. Particle Accel. Conf.} (Dresden, Germany, 2014), p. 2549.
\bibitem{AIP03Singer} W. Singer, X. Singer, J. Tiessen, H. Wen, {\em Proc. AIP Conf.} (NY, USA, 2003), p. 671.
\bibitem{IEEE09Champion} M. S. Champion, {\it et al.}, IEEE Trans. Appl. Superconductivity {\bf 19}, 1384 (2009).
\bibitem{PAC03Jiang} H. Jiang, T. R. Bieler, C. Compton, T. L. Grimm, {\em Proc. IEEE Particle Accel. Conf.} (Oregon, USA, 2003), p. 1359.
\bibitem{International07Abbas} N. M. Abbas, D. G. Solomon, Md. F. Bahari, Int. J. Machine Tools $\&$ Manufacturer, {\bf 47}, 1214 (2007).
\bibitem{RF03Bauer} P. Bauer, T. Berenc, C. Boffo, M. Foley, M. Kuchnir, Y. Tereshikin, T. Wokas,
{\em Proc. 11th Workshop on RF Superconductivity}, (Lubeck, Germany, 2003), p. 588.
\bibitem{Met09Choi} G. Choi, J. Lim, N. R. Munirathnam, I. Kim, Metals and Materials International, {\bf 15}, 385 (2009).
\bibitem{Mat00Koethe} A. Koethe, J. I. Moench, Materials Trans, {\bf 41}, 7 (2000).
\bibitem{Pfann} W. G. Pfann, {\em Zone Melting} (Wiley, New York, 1966).
\end{references}
\end{document}